\documentclass[aps,prd,floats,preprint,nofootinbib]{revtex4}
\usepackage{graphicx}
\usepackage{amsmath}
\usepackage{amsfonts}
\usepackage{amssymb}
\usepackage{multirow}
\usepackage{psfrag}
\usepackage{color}

\begin{document}
\def\beq{\begin{equation}}
\def\eeq{\end{equation}}
\def\bea{\begin{eqnarray}}
\def\eea{\end{eqnarray}}
\def\ben{\begin{enumerate}}
\def\een{\end{enumerate}}
\def\ie{{\it i.e.}}
\def\etc{{\it etc.}}
\def\eg{{\it e.g.}}
\def\lsim{\mathrel{\raise.3ex\hbox{$<$\kern-.75em\lower1ex\hbox{$\sim$}}}}
\def\gsim{\mathrel{\raise.3ex\hbox{$>$\kern-.75em\lower1ex\hbox{$\sim$}}}}
\def\ifmath#1{\relax\ifmmode #1\else $#1$\fi}
\def\half{\ifmath{{\textstyle{1 \over 2}}}}
\def\threehalf{\ifmath{{\textstyle{3 \over 2}}}}
\def\quarter{\ifmath{{\textstyle{1 \over 4}}}}
\def\eigth{\ifmath{{\textstyle{1\over 8}}}}
\def\sixth{\ifmath{{\textstyle{1 \over 6}}}}
\def\third{\ifmath{{\textstyle{1 \over 3}}}}
\def\twothirds{{\textstyle{2 \over 3}}}
\def\fivethirds{{\textstyle{5 \over 3}}}
\def\fourth{\ifmath{{\textstyle{1\over 4}}}}
\def\chitil{\wt\chi}
\def\fbi{~{\mbox{fb}^{-1}}}
\def\fb{~{\mbox{fb}}}
\def\br{BR}
\def\gev{~{\mbox{GeV}}}
\def\calm{\mathcal{M}}
\def\mll{m_{\ell^+\ell^-}}
\def\tanb{\tan\beta}

\def\wtil{\widetilde}
\def\cnone{\wt\chi^0_1}
\def\cnonestar{\wt\chi_1^{0\star}}
\def\cntwo{\wt\chi^0_2}
\def\cnthree{\wt\chi^0_3}
\def\cnfour{\wt\chi^0_4}
\def\snu{\wt\nu}
\def\snul{\wt\nu_L}
\def\msnul{m_{\snul}}
\def\se{\wt e}
\def\smu{\wt\mu}
\def\snu{\wt\nu}
\def\snul{\wt\nu_L}
\def\msnul{m_{\snul}}

\def\snue{\wt\nu_e}
\def\snuel{\wt\nu_{e\,L}}
\def\msnuel{m_{\snul}}

\def\snubar{\ov{\snu}}
\def\msnu{m_{\snu}}

\def\snue{\wt\nu_e}
\def\snuel{\wt\nu_{e\,L}}
\def\msnuel{m_{\snul}}

\def\snubar{\ov{\snu}}
\def\msnu{m_{\snu}}
\def\mcnone{m_{\cnone}}
\def\mcntwo{m_{\cntwo}}
\def\mcnthree{m_{\cnthree}}
\def\mcnfour{m_{\cnfour}}
\def\wt{\widetilde}
\def\anti{\overline}
\def\wh{\widehat}
\def\cpone{\wt \chi^+_1}
\def\cmone{\wt \chi^-_1}
\def\cpmone{\wt \chi^{\pm}_1}
\def\mcpone{m_{\cpone}}
\def\mcpmone{m_{\cpmone}}

\def\staur{\wt \tau_R}
\def\staul{\wt \tau_L}
\def\stau{\wt \tau}
\def\mstaur{m_{\staur}}
\def\stauone{\wt \tau_1}
\def\mstauone{m_{\stauone}}

\def\gl{\wt g}
\def\mgl{m_{\gl}}
\def\stl{{\wt t_L}}
\def\str{{\wt t_R}}
\def\mstl{m_{\stl}}
\def\mstr{m_{\str}}
\def\sbl{{\wt b_L}}
\def\sbr{{\wt b_R}}
\def\msbl{m_{\sbl}}
\def\msbr{m_{\sbr}}
\def\sbot{\wt b}
\def\msbot{m_{\sbot}}
\def\sq{\wt q}
\def\sqbar{\ov{\sq}}
\def\msq{m_{\sq}}
\def\slep{\wt \ell}
\def\slepbar{\ov{\slep}}
\def\mslep{m_{\slep}}
\def\slepl{\wt \ell_L}
\def\mslepl{m_{\slepl}}
\def\slepr{\wt \ell_R}
\def\mslepr{m_{\slepr}}
\def\jet{{\rm jet}}
\def\filt{{\rm filt}}
\def\cut{{\rm cut}}
\def\sub{{\rm sub}}
\def\sig{\mbox{sig}}

\def\fT{f_T}
\def\fpi{f_\pi}
\def\nsub{ n_{\text{sub}}}
\def\tbar{\overline{t}}
\def\ubar{\overline{u}}
\def\cbar{\overline{c}}
\def\Tbar{\overline{T}}
\def\bbar{\overline{b}}
\def\afb{A_{FB}^t}
\def\ttbar{t\tbar}
\def\uubar{u\ubar}
\def\tbart{\tbar t}
\def\bbbar{b\bbar}
\def\hc{ \text{h.c.}}
\def\detPhi{\text{det}~\Phi}
\def\e{\varepsilon}
\def\Oone{\mathcal{O}(1)}
\def\M{\mathcal{M}}
\definecolor{darkred}{rgb}{0.7,0.0,0.0}
\definecolor{darkblue}{rgb}{0.0,0.0,0.9}
\definecolor{darkgreen}{rgb}{0.0,0.5,0.0}
\definecolor{brown}{rgb}{0.0,0.0,0.0}
\definecolor{white}{rgb}{1.0,1.0,1.0}

\newcommand{\red}{\color{darkred}}
\newcommand{\blue}{\color{darkblue}}
\newcommand{\brown}{\color{brown}}
\newcommand{\green}{\color{darkgreen}}
\newcommand{\white}{\color{white}}
\newcommand{\excola}{\white}
\newcommand{\excolb}{\blue}

\newcommand{\gtupi}{ { g_{t u \pi}}}
\newcommand{\gtuh}{ { g_{t u h_t}}}
\newcommand{\gturho}{ { g_{t u \rho}}}

\def\CC{{C\nolinebreak[4]\hspace{-.05em}\raisebox{.4ex}{\tiny\bf ++}}}
\newcommand{ \slashchar }[1]{\setbox0=\hbox{$#1$}   
   \dimen0=\wd0                                     
   \setbox1=\hbox{/} \dimen1=\wd1                   
   \ifdim\dimen0>\dimen1                            
      \rlap{\hbox to \dimen0{\hfil/\hfil}}          
      #1                                            
   \else                                            
      \rlap{\hbox to \dimen1{\hfil$#1$\hfil}}       
      /                                             
   \fi}
\widowpenalty=1000
\clubpenalty=1000
\vspace*{3cm}
\title{Top condensation as a motivated explanation of the top forward-backward asymmetry}

\author{Yanou Cui,  Zhenyu Han, and Matthew D. Schwartz}

\affiliation{ \small \sl Center for the Fundamental Laws of Nature\\ Harvard University,\\ Cambridge, MA 02138, USA}

\def\thesection{\arabic{section}}
\def\thetable{\arabic{table}}


\begin{abstract}
Models of top condensation can provide both a compelling solution
to the hierarchy problem as well as an explanation of why the top-quark mass is large.
The spectrum of such models, in particular topcolor-assisted technicolor,  includes top-pions, top-rhos and the top-Higgs,
all of which can easily have large top-charm or top-up couplings. Large top-up couplings in particular
would lead to a top forward-backward asymmetry through $t$-channel exchange, easily consistent with
the Tevatron measurements. Intriguingly, there is destructive interference between the top-mesons
and the standard model which conspire to make the overall
top pair production rate consistent with the standard model. The rate for same-sign top production is also
small due to destructive interference between the neutral top-pion and the top-Higgs. Flavor physics is under control because new physics is mostly confined to the top quark. In this way, top condensation can explain the asymmetry and be consistent with all experimental bounds. There are many additional signatures of topcolor with large $tu$ mixing, such as top(s)+jet(s) events, in which
a top and a jet reconstruct a resonance mass, which make these models easily testable at the LHC.
\end{abstract}

\maketitle
\thispagestyle{empty}



\section{Introduction}
\label{sec:introduction}
  The CDF collaboration recently reported the measurement of a large forward-backward
asymmetry ($\afb$) in top pair production with 5~fb$^{-1}$
of data in both semi-leptonic~\cite{cdf-1} and dileptonic~\cite{cdf-2} channels. The observed asymmetry,
for $m_{\ttbar} > 450$ GeV deviates from the standard model by more than 3$\sigma$.
The new CDF results are consistent with earlier observation of large top asymmetries by both CDF and D0 based on
smaller data sets \cite{d0-2007,cdf-08,cdf-09}, and moreover
the significance of the discrepancy has grown over time. Constraints from other measurements,
such as the total $\ttbar$ cross section and the dijet bound, make model building curiously difficult.

There have been basically two classes of models proposed to explain the top quark anomaly, $s$-channel
and $t$-channel models, both of which must have non-universal flavor structure.
The first class of models provide a tree-level contribution to $\ttbar$ through the exchange of a
new particle in the $s$-channel. This contribution cannot be too large without affecting the overall $\ttbar$ rate, so the new
particle should be heavy ($\gsim$ 1 TeV). Even then, one expects the cross section to grow
as a function of the $\ttbar$ invariant mass, which
is not seen.  The contribution to $\afb$ can be enhanced by making the new particle colored and parity-violating, which
leads to axigluon models~\cite{Frampton:2009rk}. However, to get the right $\afb$ contribution,
the axigluon has to have unusual opposite sign couplings to $u\bar{u}$ and $\ttbar$. Moreover, dijet bound constraints
force these models to couple more weakly to ups than tops~\cite{Bai:2011ed}.

The $t$-channel models can explain $\afb$ if some new particle has $\Oone$ couplings to up and top~\cite{Jung:2009jz}.
$t$-channel models are not constrained by the dijet data, but there are constraints from the total $\ttbar$ rate, as well
as same-sign tops at the Tevatron, and single top production. In these models, the large flavor-violating $tu$ coupling appears initially
to be ad-hoc and unnatural.

In order to make $t$-channel models more appealing theoretically, there have been attempts to embed them in flavor-conserving models by introducing new horizontal symmetries \cite{Grinstein:2011yv, Ligeti:2011vt, Jung:2011zv, Nelson:2011us, Babu:2011yw}.
In the existing models, the top asymmetry anomaly is explained by new physics uncorrelated to the profound puzzle of electroweak
symmetry breaking (EWSB) which necessitates new physics interpretation on its own.
The models would be much more compelling if the $\afb$ anomaly corresponded to a previously
existing mechanism of EWSB. In this paper, we demonstrate that there is indeed such a possibility in the framework of top condensation \cite{Miransky:1988xi, Miransky:1989ds, Bardeen:1989ds, Hill:1991at}.
Top condensation models can provide particles which naturally have large top-up couplings, and can be exchanged in the $t$-channel
to produce the observed $\afb$.

In top condensation, electroweak symmetry is dynamically broken, as in technicolor (and in QCD),
through the formation of bound states.  The difference
from technicolor is that the condensate is made up of the top quark itself, rather than new techniquarks.
A realistic and viable framework for top condensation is topcolor assisted technicolor (TC2) \cite{Hill:1994hp}. In this framework,
there are two sources of EWSB, or three if QCD is included. The majority of EWSB and the majority of the the $W$ and $Z$ masses
are due to technicolor, with a small contribution from topcolor; the majority of the
top quark mass comes from the topcolor condensate. The light quarks can get mass from the extended technicolor or a scalar Higgs.
In this way, TC2 solves one of the difficulties of technicolor, namely how to explain large top Yukawa without flavor problem.
We will review the TC2 setup in Section~\ref{sec:TC2}.

One generic consequence of having two sources of electroweak symmetry breaking is that there are two sets of Goldstone bosons. One
set is eaten by the $W$ and $Z$ to give them their masses, the other set are an $SU(2)$ triplet of top-pions. These top-pions
are similar to the charged Higgses and pseudoscalar
 in two-Higgs doublet models, but couple only to the third generation, at leading order.
Topcolor also generically has a top-Higgs, which is the scalar $\ttbar$ bound state.
In addition, there should be angular and radial excitations.
The lightest of these, in analogy to QCD, is expected to be a vector, the top-rho. 
In the unbroken phase, all of these particles couple predominantly to $\ttbar$, where here $t$ is the top-color eigenstate.

After electroweak symmetry, the right-handed top can have large mixing with the right-handed charm and up-quarks. 
That only the right-handed top quark has large mixing is a consequence of having only a $\ttbar$ condensate, not $\bbbar$,
which is in turn a consequence of top-hypercharge being attractive in the $\ttbar$ channel but repulsive in $\bbbar$.
While there are strong constraints on left-handed mixing, there is substantial freedom in mixing $t_R$ without
violating flavor bounds, an appealing natural feature of this model.
 If there is substantial mixing between $t_R$ and $u_R$, there will
be large flavor changing  $tu$ couplings in the mass basis. Thus the top-pions, top-Higgs, and top-rho can all contribute to $\ttbar$ production,
with exactly the desired features of the $t$-channel models to explain $\afb$. 

One critical feature for the viability of this model is the large interference effects among the new physics particles,
and between the new physics and the standard model. The interference between the top-mesons and the standard model in
$\ttbar$ production is destructive, making it possible to be consistent with the measured total $\ttbar$ cross section and at the same time maintain a large enough $\afb$. Large $tu$ couplings generically predict the production of abundant same-sign top quarks. This is also true in our model. However, due to destructive interference between the top-Higgs and the top-pion, as well as that between the top-Higgs and the top-rho, there exists large parameter space within current experimental bounds.

The organization of this paper is as follows. We begin in Section~\ref{sec:TC2} with a review of the TC2 setup. This review
is based on the Hill and Simmons technicolor review~\cite{Hill+Simmons}, with a slightly different emphasis and a few more
relevant details. The relevant low energy theory after mixing is expanded in Section~\ref{sec:EFT}.  In Section~\ref{sec:pheno}, we isolate the important relevant couplings in the low energy theory and
discuss their phenomenological implications. In particular, we discuss $\afb$ in Section~\ref{sec:afb}, the $\ttbar$ rate
in Section~\ref{sec:ttbar} and the same-sign top rate in Section~\ref{sec:tt}. The bounds are combined in Section~\ref{sec:comb}.
 Section~\ref{sec:more} discusses additional constraints and signatures.
 Section~\ref{sec:conc} presents our conclusions.


\section{Topcolor assisted technicolor \label{sec:TC2}}
In top condensation, the condensate which breaks electroweak symmetry is
made up of the top quark itself, rather than new techniquarks as in technicolor.
For this to work, the top quark must not be confined in the condensate. Such behavior
is different from QCD, but not unreasonable.  Indeed, exactly such a
situation happens in a non-relativistic situation with the formation of
Cooper pairs and breaking of $U(1)_\text{EM}$ in the BCS theory of superconductivity.
The behavior is consistent with calculations in the Nambu-Jona-Lasinio (NJL) model, although these calculations
are very approximate. Realistic natural models of topcolor can only provide around $60-100$ GeV of the EWSB vev
$v_0=246$~GeV, so they must be supplemented by another EWSB sector, leading
to topcolor-assisted technicolor (TC2). In Section~\ref{sec:overview} we give
an overview of top-condensation. Section~\ref{sec:model} specifies to TC2.
The low energy effective theory of TC2 including the relevant mixing is discussed in Section~\ref{sec:EFT}.

\subsection{Overview \label{sec:overview}}
Top condensation is based on the observation that strong interactions among fermions, such as with a 4-Fermi
operator with large coefficient, can lead to bound states. These bound states can pick up a negative mass squared
through RG evolution leading to spontaneous symmetry breaking.
Top condensation observes that the top quark
is already expected to have strong interactions, since its Yukawa coupling, $y_t = \sqrt{2}m_t/v\sim 1$, so it
is natural to have strong dynamics associated with the third generation alone.
In the NJL model, where one includes only a single dimension 6 operator and only Fermion loops,
the qualitative features of top-condensation can be shown to be reasonable.
In this way the proximity of the top mass to the electroweak vev is required, and explained, in
contrast to in technicolor theories, where the large top mass is extremely challenging to explain.

The typical picture in top condensation is that the top first forms a bound state,
$H_t \sim \tbart$ at a scale $M$. In topcolor models, the scale $M$ is associated with a new force, topcolor,
which acts on the third generation quarks. If topcolor is strong, and itself Higgsed so that the topcolor
gauge bosons are massive, then the leading operator generated by integrating out the topcolor gauge bosons is
\begin{equation}
  -g^2 \tbart \frac{1}{p^2 - M^2} \tbart \to \frac{g^2}{M^2} \tbart\tbart \sim g \tbart H_t -\frac{1}{2} m_H^2 H_t^2,
\end{equation}
where the top-Higgs, $H_t=\frac{g}{M^2}\tbart$ has been integrated-in in the last step and $m_H=\frac{M}{g}$ at the scale
$\mu=M$. At this point, $H_t$ is just an auxiliary field, with no kinetic term. It picks up a kinetic term
from renormalization-group evolution and becomes dynamical below the scale $M/g$. The vanishing of the kinetic term
at $\mu=M/g$ is called a  compositeness boundary condition. 

At tree level, the mass-squared of $H_t$ is positive, but it also gets radiative corrections.
By dimensional analysis,
\begin{equation}
m_H^2 = (1-\gamma) \frac{M^2}{g^2}
\end{equation}
where $\gamma$ is an anomalous dimension (in the NJL model, $\gamma= 2 N_c (\frac{g}{4\pi})^2$).
For sufficiently large $\gamma \gsim 1$, the top-Higgs mass squared can be negative, signalling spontaneous symmetry breaking.
In order for the top-Higgs to get an expectation value $v$ lower than $M$, the anomalous dimension should be close
to its critical value. That some tuning is required is not surprising, since the Higgs mass is
getting quadratic corrections. Thus it is expected $M\sim$~TeV. If $M$ is a little larger, there is a little
hierarchy problem. For much larger $M$, the full hierarchy problem is reintroduced. 

Since the top-Higgs vev and
the top mass are generated from the same dynamics, they are related. Explicitly,
\begin{equation}
  v^2 = \gamma \: m_t^2 \ln\frac{M^2}{g^2 m_t^2} + \cdots
\end{equation}
where $\cdots$ represent the contribution from additional operators generated at the scale $M$ where topcolor is integrated out
(typically these are small corrections). Since $M\sim$ TeV by naturalness, this log is small so we expect $v< m_t$. Typically
$v\sim 60$ GeV. Thus, topcolor as an explanation of electroweak symmetry breaking is incomplete.

Realistic models of top condensation have two sources of electroweak symmetry breaking: the first, topcolor, generates
a EWSB vev around 60-100 GeV. In topcolor-assisted technicolor, this is supplemented by another source of EWSB
providing the remaining contribution to $v$, such as technicolor.
Technicolor itself is a beautiful explanation
of the $W$ and $Z$ masses, but does not explain fermion masses.
Extended technicolor (ETC) can explain the fermion masses through a new force, but has flavor problems. These flavor
problems are ameliorated by assuming non-QCD like dynamics (walking technicolor) for the ETC sector in which operators
involving techniquarks run faster than operators involving standard model quarks.
However, even these models have difficulty explaining the large top quark mass. In this way, topcolor ``assists'' technicolor
by handling the top mass. The main ingredient of top-color assisted technicolor
 relevant to the top forward-backward asymmetry is simply
that there are two EWSB sectors, and one of them couples only to the top quark at leading order. Since the details of
the technicolor side are irrelevant, we will abbreviate that sector with a simple scalar Higgs doublet $H$.
This approximation is called Bosonic topcolor~\cite{Samuel:1990dq}.


\subsection{The model}
\label{sec:model}
In this section, we briefly review the relevant ingredients in a realistic topcolor-assisted technicolor model. As explained
in Section~\ref{sec:overview}, these models have two sources of electroweak symmetry breaking: one from the topcolor sector,
generating a vev $v\sim 60$ GeV, and one from another sector. For simplicity, we will take this other sector to be a single
Higgs doublet model. This can be replaced with an extended technicolor sector with little consequence for
the top asymmetry and the model bounds. Our presentation will closely follow the Hill-Simmons review~\cite{Hill+Simmons},
with a few added details and simplifications.

\begin{figure}[t]
\begin{center}
\psfrag{A}[b]{\large{{\red $SU(3)_1 \times U(1)_1$}}}
\psfrag{B}[b]{\large{{\blue $SU(3)_2 \times U(1)_2$}}}
\psfrag{C}{\large{$\Phi$}}
\psfrag{D}[b]{{\red $3^{\text{rd}}$ generation}}
\psfrag{E}[b]{{\blue $1^{\text{st}}$ and $2^{\text{nd}}$ generations}}
\includegraphics[width=0.9\textwidth]{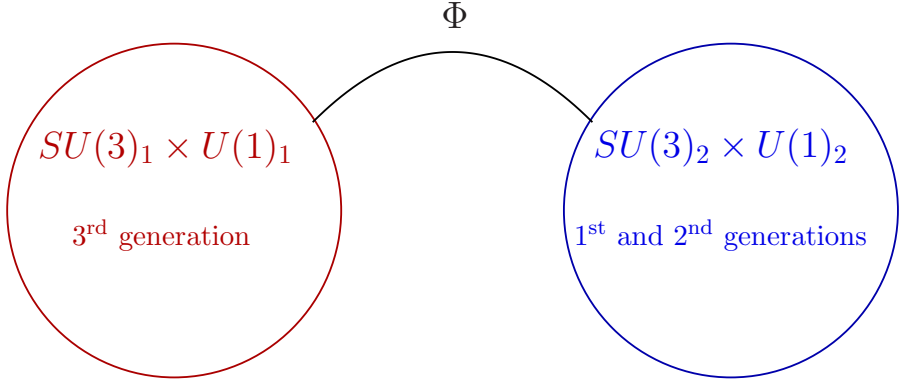}
\end{center}
\vspace{-1cm}
\caption{Moose diagram for minimal topcolor. \label{fig:moose}}
\end{figure}

\begin{table}[t]
\begin{center}
  \begin{tabular}{|c|c|c|c|c|c|}
 \hline
field & {\red $SU(3)_1$} &{\red  $U(1)_1$} &{\blue $SU(3)_2$ }& {\blue $U(1)_2$ }& $SU(2)_L$ \\
\hline
{\red $T_L$} & {\red  $\square$} & {\red $\frac{1}{3}$} & -  &  - & $\square$ \\
{\red $t_R$} & {\red  $\square$} & {\red $\frac{4}{3}$} & -  &  - &  -  \\
{\red $b_R$} & {\red  $\square$} & {\red -$\frac{2}{3}$} & -  &  - &  - \\
\hline
{\blue $C_L, U_L $} & - & - &{\blue $\square$} &{\blue $\frac{1}{3}$} & $\square$ \\
{\blue $c_R, u_R $} & - & - &{\blue $\square$} &{\blue $\frac{4}{3}$} &  -  \\
{\blue $s_R, d_R $} & - & - &{\blue $\square$} &{\blue  -$\frac{2}{3}$} &  - \\
\hline
$\Phi$ &  $\square$ & $\frac{1}{3}$ &  $\overline\square$ & $-\frac{1}{3}$ &  - \\
\hline
\hline
$\detPhi$ & - &  $1$ &  - & -1 &  - \\
\hline
  \end{tabular}
\hspace{0.2cm}
  \begin{tabular}{|c|c|c|c|c|c|}
 \hline
field & {\red $SU(3)_1$} &{\red  $U(1)_1$} &{\blue $SU(3)_2$ }& {\blue $U(1)_2$ }& $SU(2)_L$ \\
\hline
{\red $\left(\begin{array}{c}\tau_L\\ \nu_\tau\end{array}\right)$} & - &{\red-1 }& -  &  - & $\square$ \\
{\red $\tau_R$} &  - &{\red-2}& -  &  - &  -  \\
\hline
{\blue$\left(\begin{array}{c}\ell_L\\ \nu_\ell\end{array}\right)$} &  - & - & - &{\blue -1} & $\square$ \\
{\blue$\mu_R, e_R $ }& - & - & - &{\blue -2} &  -  \\
\hline
{\blue$H$} &  - & - & - &{\blue $-1$} & $\square$ \\
\hline
\hline
{\red $H_t$} &  - &{\red  $-1$} & - & - & $\square$ \\
\hline
  \end{tabular}
\end{center}
\caption{Particle content and quantum numbers fields in minimal TC2. The fields $\detPhi$ and $H_t$ (and $H$ in technicolor)
are composite
and useful for writing down the effective Lagrangian. \label{tab:content}}
\end{table}

In minimal topcolor, the third generation couples to one copy of $SU(3)\times U(1)$ gauge group and the first two generations to another
copy of $SU(3)\times U(1)$ gauge group.  The standard model $SU(3)_\text{QCD} \times U(1)_Y$ is the diagonal combination of these two groups.
The  $SU(3)$ group coupling to the 3rd generation (topcolor) must be much stronger than the other $SU(3)$, to generate the
top condensate. The two $U(1)$'s are necessary so that only a $\langle \ttbar \rangle$ forms and not $\langle b\bar{b} \rangle$.
In this setup, there is one $SU(2)_\text{weak}$ which couples to everything. Variations are possible, with 3 copies
of $SU(3)\times SU(2)\times U(1)$, one for each generation (which mimic extra-dimensional models), or the top-triangle
moose~\cite{Chivukula:2009ck} (inspired by Higgsless models), which has two $SU(2)$'s, but we stick to this minimal construction for simplicity.

In addition to the standard model and the new gauge bosons for the extra $SU(3)$ and $U(1)$, we need a field $\Phi$ which
spontaneously breaks topcolor, giving the topgluons a mass $M$ of order $\sim1$ TeV. When $\Phi$  gets a vev,
\begin{equation}
SU(3)_1\times SU(3)_2\times U(1)_1\times U(1)_2
\overset{ \langle \Phi \rangle}{\to} SU(3)_\text{QCD} \times U(1)_Y.
\end{equation}
So $\Phi$ should be charged under all these groups. In particular, if $\Phi$ is a bifundamental, as shown
in the moose diagram in Figure~\ref{fig:moose}, it will automatically
break the group down to the diagonal.
We also include a Higgs $H$ which couples
to everything to parameterize the technicolor contribution to EWSB.
There are 4 couplings $g_1^s, g_2^s, g^Y_1$ and $g^Y_2$ associated with the two $SU(3)$'s and two $U(1)$'s.
The field content is given in Table~\ref{tab:content}.

In top-condensation, when $H_t = \ttbar$ gets a vev electroweak symmetry will be broken.
Since the bottom quark is charged also under $SU(3)_1$,
one must be careful not to have $\langle \bbbar \rangle$ as well, which would generate
a large bottom quark mass. The $U(1)_1$ hypercharge conveniently can achieve this.
Indeed, with the standard model hypercharge assignments, this $U(1)$ is attractive in the $\ttbar$ channel and
repulsive in the $\bbbar$ channel. If the coupling is large (but not too large or else $\langle
\tau \bar{\tau} \rangle \ne 0$), this can allow for only a $\ttbar$ condensate~\cite{Hill+Simmons}.

In TC2, electroweak symmetry is broken by both top condensation, through $\langle H_t \rangle = \langle \ttbar \rangle = f_\pi$
and by technicolor through $\langle H \rangle=f_T$. With two sources of EWSB, there will be two sets
of Goldstone bosons. One set are eaten by the $W$ and $Z$, and the other set are called top-pions.

\subsection{Effective Theory \label{sec:EFT}}
The easiest way to study the phenomenology of this model is through an effective linear sigma model.
To do this we restore the full gauge symmetries by introducing a sigma field
$\Sigma= \exp(i \pi^a \tau^a/\sqrt{2}f_\pi)$, where $\tau_a$ are the Pauli matrices.
The top-Higgs doublet in the linear sigma model can be represented by
\begin{equation}
  H_t = \Sigma
  \begin{pmatrix}
f_\pi +\frac{1}{\sqrt{2}} h_t\\0
  \end{pmatrix}
=
  \begin{pmatrix}
f_\pi+\frac{1}{\sqrt{2}}( h_t + i \pi^0)\\
i \pi^+
  \end{pmatrix}
+\cdots
\end{equation}
With this normalization a kinetic term $(D_\mu H)^\dag(D_\mu H)$ gives the proper normalization to the top-pions and
the top-Higgs. In the chiral Lagrangian for the top-pions, the kinetic term is normalized to
$\frac{\fpi^2}{2} \text{Tr}[(D_\mu \Sigma)^\dag (D_\mu \Sigma)]$. In TC2, there is a second Higgs, $H$, which contains the pions
from technicolor with decay constant $\fT$, with similar kinetic terms. The electroweak vev
$v_\text{ew}=\frac{v_0}{\sqrt{2}}=175 \gev$ gets
contributions from both $v_\text{ew}^2 = \fpi^2 +\fT^2$. The expectation from the NJL model is that $f_\pi\sim 60$ GeV, although
$\fpi$ can really be anything. Constraints from $Z\to \bbbar$ suggests that $\fpi \gsim 100$ GeV~\cite{Burdman:1997pf}.

Including the the Higgs fields $H_t$ and $H$ and the link field $\Phi$ restores the full gauge symmetry.
Then, including the operators allowed by the full symmetries, we can construct the effective Lagrangian.
First, the top-pion interactions are generated from the top mass term via
\begin{align}
  \mathcal{L}_\Sigma &= \lambda_t\Tbar_L H_t t_R +\hc\\
&= \frac{\lambda_t f_\pi}{\sqrt{2}} \ttbar
+ \frac{\lambda_t}{\sqrt{2}}\left( i \pi^0 \tbar \gamma^5 t+ h_t \ttbar\right)+ \lambda_t \pi^+ \tbar_R b_L + \lambda_t \pi^- \bbar_L t_R + \cdots \label{eq:piterms}
\end{align}
Here we can identify $m_0 = \frac{\lambda_t\fpi}{\sqrt{2}}$ as the topcolor
contribution to the top mass, and the top-pion couplings, before diagonalization, as $g_{\pi t t} = \frac{m_0}{\fpi}$
and $g_{\pi b t} = \sqrt{2} g_{\pi t t}$. The relationship between the top-Higgs and top-pion couplings here is particular
to the linear-sigma model and not constrained by symmetries.~\footnote{In some sense,
a linear multiplet is more natural in topcolor than in technicolor. 
Since the topcolor gauge coupling must be close to its critical value, the spontaneous breaking of chiral symmetry is
a relatively small effect, $f_t \ll M$. Close to criticality, the  phase transition should be smooth, and
the linear representation of bound state $H_t$ should be a fairly good transition even after chiral symmetry is broken~\cite{Chivukula:1998uf}.}

The (technicolor) Higgs field has normal Yukawa interactions with the first two generations
\begin{align}
  \mathcal{L}_H &= Y^u_{22} C_L H c_R  + Y^u_{12} C_L H u_R + Y^u_{21} U_L H c_R + Y^u_{11} U_L H u_R \nonumber \\
                &+ Y^d_{22} C_L H^\dag s_R  + Y^d_{12} C_L H d_R + Y^d_{21} U_L H s_R + Y^d_{11} U_L H d_R + \hc.
\end{align}
Note that light quark masses can be generated in technicolor with much less walking than heavy quark masses,
so this model alleviates a lot of the tension in ETC models.

Finally, there are terms allowed by symmetry which mix the third generation with
the other two. These must involve both $\Phi$ and $H$ to be invariant under the $U(1)$'s. These operators
must be at least dimension 5, and we characterize them with a scale $\Lambda$. The leading operators
relevant to the subsequent discussion are
\begin{multline}
  \mathcal{L}_{H\Phi} =
  c_{32} \overline T_LHc_R\frac{\Phi}{\Lambda}
+ c_{31} \overline T_LHu_R\frac{\Phi}{\Lambda}
+ d_{32} \overline T_LH^\dag s_R\frac{\Phi}{\Lambda}
+ d_{31} \overline T_LH^\dag d_R\frac{\Phi}{\Lambda}
\\
+ c_{23} \overline{C}_L H t_R\frac{\Phi^\dag}{\Lambda}\frac{\detPhi^\dag}{\Lambda^3}
+ c_{13} \overline{U}_L H t_R\frac{\Phi^\dag}{\Lambda}\frac{\detPhi^\dag}{\Lambda^3}
+ d_{23} \overline{C}_L H^\dag b_R\frac{\Phi}{\Lambda}\frac{\detPhi}{\Lambda^3}
+ d_{13} \overline{U}_L H^\dag b_R\frac{\Phi}{\Lambda}\frac{\detPhi}{\Lambda^3} \\
+ c_{33} \Tbar_L t_R H \frac{\detPhi}{\Lambda^3} + d_{33} \Tbar_L H^\dag \frac{\detPhi^\dag}{\Lambda^3}.
\label{eq:cds}
\end{multline}
The $\detPhi$ factors are necessary to maintain the $U(1)_1\times U(1)_2$ symmetry.

The scale $\Lambda$ should be larger than the $\Phi$ vev.
That is,  $\Lambda > \langle \Phi \rangle = M$, where $M$ is the topgluon mass. Since the topcolor
interactions are strong, we expect $\e \equiv M/\Lambda \lsim 1$.
Once topcolor and electroweak symmetry are broken, by
 $\langle H \rangle = \fT$ and $\langle H_t \rangle = \fpi$,
 the effective low energy quark mass matrices are
\begin{equation}
M_U=f_T
\begin{pmatrix}
Y^u_{11}    & Y^u_{12}  & c_{13} \e^4\\
Y^u_{21}    & Y^u_{22}  & c_{23} \e^4\\
c_{31} \e  & c_{32}\e &~~~ \frac{m_0}{f_T} + c_{33} \e^3,
\end{pmatrix}
\quad
M_D = f_T
\begin{pmatrix}
Y^d_{11}    & Y^d_{12}  & d_{13} \e^4\\
Y^d_{21}    & Y^d_{22}  & d_{23} \e^4\\
d_{31} \e  & d_{32}\e  & d_{33} \e^3.
\label{eq:massmatrix}
\end{pmatrix}
\end{equation}
For the top quark mass to be generated by topcolor and the bottom quark mass not to be too small, we expect
$\e \lsim 0.3$ \cite{He:1998ie, Burdman:1999sr}.

Actually, $\e$ can be much smaller \cite{Burdman:2000in},  since there is another contribution to the bottom
mass due to non-perturbative effects. 
Scaling arguments predict that $m_b \sim \frac{3}{8 \pi^2} m_t = 6.6$ GeV, which can be interpreted
as an instanton effect~\cite{Hill:1994hp}.
 Alternatively, $\e$ can even be $\Oone$, if $d_{33}$ and $c_{33}$ are very small.
For $\e\sim 1$ there are many more relevant higher dimension operators and the theory is not predictive.
We will suppose that $\e \sim 0.3$ and $c_{23}, c_{13} \sim 3$ so that there will be $\Oone$ mixing between the right-handed
top and the first two generations. It is this mixing which generates a large $\afb$.

To proceed, we will assume that terms of order $\e^3$ or higher can be neglected,
but the other terms are free parameters. Considering the possible instanton contribution, we
 allow for the $b$-quark mass to be arbitrary as well. Then the effective mass matrices reduce to
\begin{equation}
M_U= \fT
\begin{pmatrix}
Y^u_{11}    & Y^u_{12}  & 0 \\
Y^u_{21}    & Y^u_{22}  & 0 \\
Y^u_{31}    & Y^u_{32}  & \frac{m_t}{f_T} \\
\end{pmatrix},
\quad
M_D = \fT
\begin{pmatrix}
Y^d_{11}    & Y^d_{12}  & 0\\
Y^d_{21}    & Y^d_{22}  & 0\\
Y^d_{31}    & Y^d_{32}  & \frac{m_b}{\fT}
\label{eq:massmatrixtwo}.
\end{pmatrix}
\end{equation}
For the working point $\fpi=100$ GeV we have $f_T = 144$ GeV using $\fpi^2 + \fT^2 = (175 \gev)^2$. So $m_t/f_T =1.2$. Thus there can
easily be $\Oone$ mixing between the right-handed top and first two generation up-type quarks. 

The one prediction of
topcolor here is the zeros in the 3rd column.
The masses are diagonalized with rotations $M_U = U_L M_U^\text{diag} U_R^\dagger$ and $M_D  = D_L M_D^\text{diag} D_R^\dagger$,
constrained by the CKM matrix $K=U_L D_L^\dag$. As discussed in Ref.~\cite{Buchalla:1995dp}, the zeros in Eq.~(\ref{eq:massmatrixtwo}) imply $U_L^{i,3}$ and $U_L^{3,i}$ $(i\neq3)$ are almost vanishing. On the other hand, $U_R^{i,3}$ and $U_R^{3,i}$ can be large when the couplings $Y^u_{31}$ and $Y^u_{32}$ are large. This means the right-handed top can have large mixing with the first two generations.
Going to the mass basis rotates the right-handed top quark as
\begin{equation}
  t_R \to U^R_{11} t_R + U^R_{12} c_R + U^R_{13} u_R.
\end{equation}
This generates large flavor changing top couplings from the couplings in Eq.~\eqref{eq:piterms}. We find
\begin{align}
  \mathcal{L}_\pi  & =i g_{tt\pi} ( \pi^0 \tbar \gamma^5 t) +i g_{t u \pi} ( \pi^0 \tbar_L u_R) + i g_{t c \pi}( \pi^0 \tbar_L c_R) \nonumber\\
                  & ~~~~~~~~~~
+ g_{t t h_t} ( h_t \tbar t) + g_{t u h} (h_t \tbar_L  u_R) + g_{t c h_t}( h_t \tbar_L c_R) + \hc,
\end{align}
where $g_{i j\, \pi} =  m_t/\fpi U^R_{ij} \le m_t/f_\pi$.
In the linear sigma model, the top-Higgs couplings are the same as the neutral top-pion couplings. Indeed the two fields
form a complex scalar. This relation may not hold in a more complete theory, but it is a reasonable starting point and
simplifies the parameter space.

The charged pion couplings are related to the
neutral pion couplings by symmetry, as in Eq.~\eqref{eq:piterms}.
For example, the interaction $\pi^- \bbar_L t_R$ has coupling $g= U^R_{33} m_t/\fpi$. 

Topcolor models in general should also contain another $SU(2)$ doublet, the bottom-Higgs $H_b = \bar{b}_R T_L$.
This bound state should be present in the effective 
Lagrangian~\cite{Chivukula:1998uf} and may be light (of order the top-higgs mass).
We will not discuss the bottom-Higgs further simply because it is unrelated to $\afb$.
 
Finally, there are the couplings of various excitations of the $\ttbar$ condensate. The lightest
excitation is expected to be the top-rho. The top-rho couplings can be modeled, such as using
hidden local symmetry models or vector meson dominance. We simply assume that before mixing the top-rho couples
to $\ttbar$ with a strength that is a free parameter. After mixing, it picks up couplings to right-handed
up and charm quarks, just like the top-pions and the top-Higgses. The top-rho is expected to be heavier than the top-pion and the top-Higgs \cite{Chivukula:1990bc}, but have to be moderately light ($\lsim 1$ TeV) to give sizable $\afb$.  For completeness, we write its Lagrangian as
\begin{equation}
  \mathcal{L}_\rho   =
g_{tt\rho}\, \rho_\mu \tbar \gamma^\mu t
 + g_{t c\rho}\, \rho_\mu \tbar_R \gamma^\mu c_R
+ g_{t u\rho}\, \rho_\mu \tbar_R \gamma^\mu u_R.
\end{equation}
We will only discuss the top-Higgs, top-pions and top-rho in the phenomenology section.

There are additional fields in TC2, such as the bottom-Higgs, the charged rho's, the topgluon, the excitations of the link field $\Phi$,
and all the regular ETC particles. For constraints
and signatures of these other particles, we refer the reader to the review~\cite{Hill+Simmons}. None of these
fields are relevant to $\afb$, so we will ignore them.

\subsection{Flavor and electroweak constraints}
Before moving on to phenomenology of topcolor model with large flavor-violating associated with $t_R$, we would like to briefly discuss related flavor constraints and how such model could be compatible with them. 

There is no GIM mechanism in TC2 and hence there is a real danger
of generating too large FCNCs, for example, through top-pion exchange. However, the top-pions only couple to the third generation
before mixing. Large flavor-changing couplings in the first two generations are only produced
 when {\it both} the right-handed and the left-handed tops have large mixings with the first two generations. This is avoided because only the right-handed top has large mixings as we have discussed. 
Flavor constraints involving the top are very weak, so the only concern left is flavor problems coming from the $b$-sector.
In fact, there are in strong $b$-physics constraints on FCNC operators involving $t_L$ and $u_R$ and the $Z$
because these produce tree-level FCNCs involving $b_L$ after EWSB~\cite{Fox:2007in}. In our case, 
these bounds do not apply because the top-mesons decay hadronically and there are no electroweak penguins
involving the neutral top-mesons alone. Flavor constraints in next-to-minimal flavor violating
models~\cite{Agashe:2005hk}, in which new physics couples only to the third generation, are in general fairly weak.

There are additional constraints from the down-type sector, which is not
directly related to the flavor-violation necessary for $\afb$. Even in the down sector, the constraints
will be absent if the $Y^d_{31}$ and $Y^d_{32}$ elements are small.
These terms originate from the $d_{31}$ and $d_{32}$ terms in Eq.~\eqref{eq:cds} which involve $d_R$ and $s_R$. It is easy
to believe that these terms might be small for a reason related to the smallness of the down and strange quark masses.
In any case, that $d_{31}$ and $d_{32}$ are small is a standard assumption in TC2 and we have no further insight into its dynamical
origin. Our observation here is simply that $c_{31}$ or $c_{32}$ do not have to be small as well.

Based on the reasonable assumption that down-sector RH and LH mixings are both small, it is easy to see that potential dangerous $B-\bar{B}$ mixing induced via e.g. tree-level exchanging an s-channel $\pi^0$ or via box diagram involving $t_R,\pi^{-}$ can be efficiently suppressed.  Meanwhile, there is concern from $D-\bar{D}$ mixing which involves the up-sector only, and may get large contribution from exchanging $\pi^0$ if both $t_R-u_R$ and $t_R-c_R$ are sizable. One way to relieve this while keeping possible large flavor violation is to have $c_{31}$ or $c_{32}$ small, but not both. For our later discussion we will assume a large $c_{31}$. 

The model may also receive important electroweak constraint from $Z\rightarrow \bar b b$ since it directly involves the unsuppressed  $\pi^- \bbar_L t_R$ vertex in the loop \cite{Burdman:1997pf,Braam:2007pm}. However, this constraint is subject to large uncertainties in subleading calculations \cite{Hill+Simmons}. Moreover, according to related studies in \cite{Hill:1994di,Burdman:1997pf,Hill+Simmons} vector states such as top-rhos may cancel the contribution from the top-pion if they are moderately light ($\lesssim600\rm GeV$), which is also what we need to generate significant $A_{FB}^t$, as discussed later. The charged top-pions could also be heavy, which would further alleviate $R_b$.


\section{Phenomenology}
\label{sec:pheno}
In the previous section we reviewed some features of topcolor-assisted technicolor. This explanation of electroweak symmetry naturally
explains both the hierarchy and the large top Yukawa with fewer flavor problems than technicolor has without topcolor's help.
Unfortunately, very little of this model's spectrum is currently calculable. However, some general features were noted
\begin{itemize}
\item Electroweak symmetry is broken by both technicolor and topcolor. So in addition to the Goldstone bosons eaten by
the $W$ and $Z$, there is another set, the top-pions, which are $\ttbar$ bound states and predominantly couple to the top quark.
\item In addition, there is expected to be a radial excitation of $\ttbar$, the top-Higgs, and angular excitations, such as a top-rho.
\item Power counting in the low-energy effective theory implies that there can be large mixing of $t_R$ with $u_R$. This generates large flavor-changing interactions mediated by the top-mesons.
\end{itemize}
 There are other particles in the theory,
such as the charged top-pions and top-rhos, the top-gluon and all the technicolor excitations, such as the techni-rhos.
Some of these particles were discussed briefly in the previous section, and their constraints have been
discussed elsewhere. They do not contribute
directly to interesting top-quark signatures, so we will not discuss them any further.

In this section, we will discuss the effect of having large flavor changing $t u$ interactions, coming from
\begin{equation}
  \mathcal{L} =  g_\pi ( i \bar{t}_L u_R \pi^0 + \bar{t}_L u_R h_t) + g_\rho (\bar{t}_R \gamma^\mu u_R \rho_\mu) + \hc,
\label{gpirho}
\end{equation}
where we have set the top-Higgs and top-pion couplings equal to a constant we call $g_\pi$.
This does not have to be true, but it is motivated
by a linear-sigma model embedding, in which the top-pion and top-Higgs form a complex scalar $\phi = h_t + i \pi$,
and simplifies the parameter space.
There are of course plenty of other interactions of these fields. But it is these specific terms which are relevant to the
top forward-backward asymmetry.

The top-pion and top-Higgs masses are not expected to be the same, however, with the top-Higgs typically having $m_{h_t} \sim 2 m_t$ while
the top-pions, being pseudogoldstone bosons can be lighter, $m_\pi \sim  100-400 \gev$. The top-rho is expected to be heavier than both the top-pion and the top-Higgs,
based on extrapolations form QCD.
 None of these masses are calculable,
although the pion mass is often estimated using the a fermion loop approximation and the Pagels-Stokar formula \cite{Miransky:1988xi, Miransky:1989ds, Bardeen:1989ds, Hill:1991at}. In this paper,
we simply take them to be free parameters.

In this section, we describe how the top-particles can produce the measured $\afb$ while
maintaining an acceptable $\ttbar$ production rate. We will also discuss the closely related direct measurements of
same-sign tops.

\subsection{Top forward-backward asymmetry $\afb$}
\label{sec:afb}
The idea that the top forward-backward asymmetry could be evidence of a new particle, $X$, being exchanged in the $t$-channel
was proposed in \cite{Jung:2009jz}. The $t$-channel exchange works by exploiting the Rutherford singularity which enhances scattering
in the forward region. A vector boson with moderate mass, such as a $Z'$ \cite{Jung:2009jz} (or a top-rho) can generate a sizable asymmetry.
Alternatively $X$ could be a scalar \cite{Shu:2009xf,Nelson:2011us} (such as a top-Higgs or top-pion).
In the case of a scalar there is a competition between Rutherford singularity and spin-conservation,
which reduces the efficiency of generating a large $\afb$. This pushes the scalar mass towards lower end as $m_X \lesssim 200\gev$
 in order to generate $\afb \sim O(0.1)$. Both the vector and scalar by themselves will generate an excess of $\ttbar$ events.
However, if both are included, there can be destructive interference, allowing a large $\afb$ but a small effect on the $\sigma_{\ttbar}$.

The experimental measurement of $\afb$ is presented in a number of ways: in the $\ttbar$ rest frame, or in the lab frame,
and folded or unfolded, if an attempt is made to correct back to the $\ttbar$ parton level. Unfortunately the unfolding
depends on the angular distribution of the tops, which is model dependent, and unfolding using the standard model, which has a tiny
$\afb$ may give an unreliable result. In this paper, we follow the approach of~\cite{Nelson:2011us}. We impose rapidity cuts on the tops,  $|\eta_{t,\bar{t}}|<2.0$, and demand that $M_{\ttbar}>450 \gev$ and compare the $\afb$
after these cuts with reconstructed asymmetry measured by CDF (for $M_{\ttbar}>450 \gev$): $\afb = 0.210\pm0.049$.
We require the asymmetry to be within $2\sigma$ of the central value, which is about $0.1\leq \afb \leq0.3$. Our calculations are performed with Calchep v3.0 \cite{calchep} and checked with Madgraph v4.4.26~\cite{Alwall:2007st} and by hand. We use CTEQ6l PDF and choose the factorization scale and renormalization scale to be $m_t$. 

\begin{figure}
\psfrag{X}{Mass}
\psfrag{Y}{$\afb$}
\psfrag{C}{combined}
\begin{center}
\includegraphics[width=0.7\textwidth]{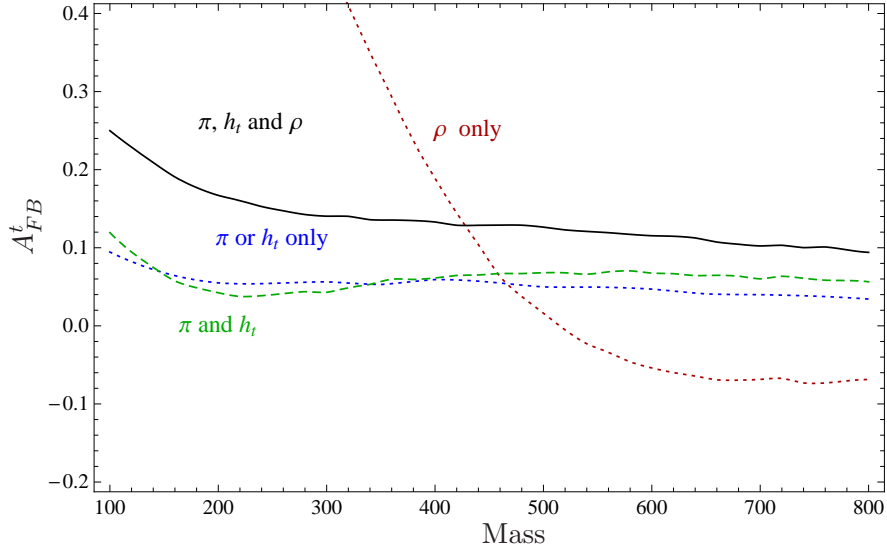}
\caption{Contribution to $\afb$ as a function of particle mass.
The blue dotted curve is from the neutral top-pion (or top-Higgs) only and the dotted red curve if for top-rho only, as functions
of the relevant particle masses. The dashed green curve has a top-pion and a top-Higgs, with $m_{h_t} =  m_\pi+ 100 \gev$
as a function of $m_\pi$.
The solid black curve also includes the top-rho, with  $m_\rho = 500 \gev$, still holding
 $m_{h_t} =  m_\pi+ 100 \gev$ as a function of $m_\pi$. All  couplings are set to 1.
\label{fig:Afb}
}
\end{center}
\end{figure}

In Figure~\ref{fig:Afb} we show the contributions of the top-pion, top-Higgs and top-rho to $\afb$ as a function
of mass, with $g_\pi=g_\rho = 1$. The single particle contributions are shown as dotted lines, with the top-pion and top-Higgs having an
identical effect. The dashed green line shows the effect of having a top-pion and a top-Higgs with $m_{h_t} = m_\pi + 100 \gev$.
The solid black curve shows the effect of adding a 500 GeV top-rho to the mix. Many
other combined curves are possible. This illustrates that even with a heavy vector, there can be a positive contribution to $\afb$
due to interference with the scalars. To get $\afb =0.2$ there is a large region of parameter space. The space become somewhat
more restricted with the $\ttbar$ cross-section and the same-sign top bounds are also imposed.

One could also consider the charged top-pion contribution to $\afb$.
The charged pions couple initially to $t_R b_L$. Due to the smallness of the $b$-quark PDF,
the $t$-channel exchange of $\pi^\pm_t$ through this interaction has a negligible contribution to $A_{FB}^t$.
$t_R-u_R$ mixing will only weaken the effect. One could imagine that there might be large $b_L-d_L$ mixing,
which would allow for $\afb$ to be created through $d\bar{d}\rightarrow t\bar{t}$ with a $t$-channel $\pi_t^\pm$.
However, $b_L-d_L$ mixing is strongly constrained by CKM unitarity bounds~\cite{Hill+Simmons},
so this contribution cannot generate a sizable asymmetry either.


\subsection{$\ttbar$ cross section}
\label{sec:ttbar}
   Introducing new processes involving new light states with sizable coupling raises concerns about large deviations in
the $\ttbar$ cross section, $\sigma_{\ttbar}$.
As pointed out in previous work e.g.\cite{Shu:2009xf}, contributions from the $t$-channel exchange amplitude squared,
which is positive, could be cancelled by a negative contribution from interference with QCD, rendering
small the net deviation from the measured $\sigma_{\ttbar}$.
Indeed there can easily be interference between different new physics contributions as well and one has to do a careful analysis.
In our model, we see that the interferences between QCD and all three new particles are negative and those among the new particles are all positive. This is easily seen by going to the limit of the $\ttbar$ threshold and examine different contributions to the matrix element squared, as shown in Table~\ref{tab:ttbar_int}. For reference, the full cross section is given in Appendix~\ref{app:ttbar}.
\begin{table}[t]
  \begin{tabular}{|c|c|c|c|c|}
    \hline
    $|{\mathcal M}(\uubar \to \ttbar)|^2$ & QCD & $\pi$ & $h_t$ & $\rho$\\
    \hline
    QCD & +$\frac{4 g_s^4}{9}$ & $-\frac{8 g_{\pi }^2 g_s^2 m_t^2}{9 \left(m_{\pi }^2+m_t^2\right)}$ & $-\frac{8 g_h^2 g_s^2 m_t^2}{9 \left(m_h^2+h_t^2\right)}$
    & $-\frac{16 g_s^2 g_{\rho }^2 m_t^2}{9 m_{\rho }^2}$\\
    \hline
    $\pi$ &  & +$\frac{g_{\pi }^4 m_t^4}{\left(m_{\pi }^2+m_t^2\right){}^2}$ &
    +$\frac{2 g_h^2 g_{\pi }^2 m_t^4}{\left(m_h^2+m_t^2\right) \left(m_{\pi }^2+m_t^2\right)}$ & 
+$\frac{4g_{\pi }^2 g_{\rho }^2  m_t^4}{ \left(m_{\pi }^2+m_t^2\right) m_{\rho }^2 }$\\
    \hline
    $h_t$ &  & &$+\frac{g_h^4 m_t^4}{(m_h^2+m_t^2)^2} $&
 +$\frac{4g_h^2 g_{\rho }^2 m_t^4 }{ \left(m_h^2+m_t^2\right) m_{\rho }^2}$\\
    \hline
    $\rho$ &  &  &  & 
+$\frac{4 g_{\rho }^4 m_t^4}{ m_{\rho }^4}$\\
    \hline
  \end{tabular}
  \caption{\label{tab:ttbar_int}
Interference terms in $t - \bar{t}$ production. To see the rough scaling properties
and signs, we have taken $s = 4 m_t^2, t =u=-m_t^2$ and $m_\rho \gg m_t$. The full matrix elements are given in Appendix \ref{app:ttbar}.
}
\end{table}   

The $\ttbar$ cross section has been measured at the Tevatron and the LHC and is in
good agreement with the standard model. There is also data on the differential $\ttbar$ cross section as a function of the $\ttbar$
invariant mass, but the error bars are still large. Thus, we compare only to the inclusive $\ttbar$ cross-section. The cross
section is enhanced at NLO at both the Tevatron and the LHC. 
The $K$-factor can depend on the experimental cuts used, so it is difficult to compare to
the measured cross section directly. Instead, we will simply require that the effect of the tree-level new physics
contributions to $\sigma_{\ttbar}$  produce less than a $20\%$ change from the standard model $\sigma_{\ttbar}$, also
computed at tree level.
This roughly corresponds to $2\sigma$ band by $\sigma_{t\bar{t}}$ measurement \cite{cdf-ttbar}.

\begin{figure}[t]
\psfrag{X}{Mass}
\psfrag{Y}{Change in $\sigma_{\ttbar}$ (\%)}
\psfrag{C}{combined}
\begin{center}
\includegraphics[width=0.7\textwidth]{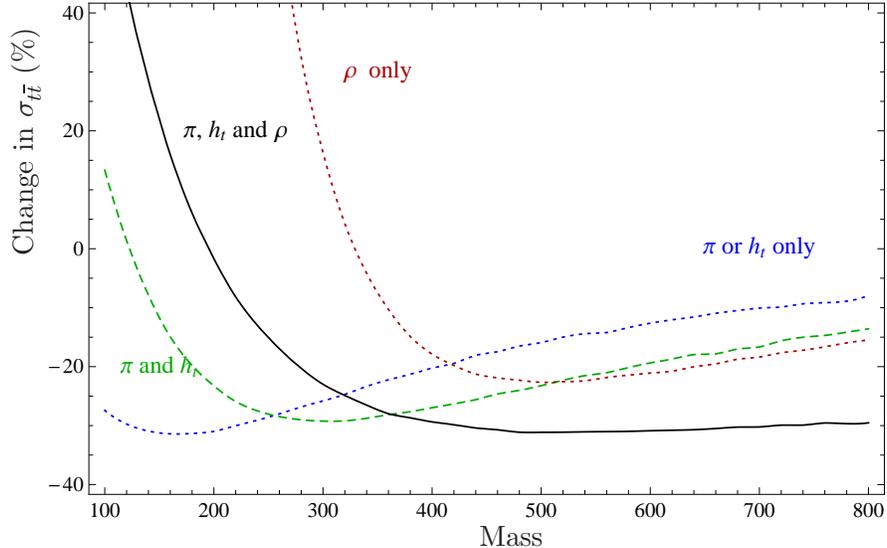}
\caption{Percent change in $\sigma_{\ttbar}$ due to new physics and interference between new physics
and the standard model, as a function of particle mass.
The blue dotted curve is from the neutral top-pion (or top-Higgs) only and the dotted red curve if for top-rho only, as functions
of the relevant particle masses. The dashed green curve has a top-pion and a top-Higgs, with $m_{h_t} =  m_\pi+ 100 \gev$
as a function of $m_\pi$.
The solid black curve also includes the top-rho, with  $m_\rho = 500 \gev$, still holding
 $m_{h_t} =  m_\pi+ 100 \gev$ as a function of $m_\pi$. All  couplings are set to 1.
\label{fig:TTsingle}
}
\end{center}
\end{figure}

Figure~\ref{fig:TTsingle} shows the percent change in the total $\ttbar$ cross section as a function of particle masses. We see that the
top-pions (or top-Higgs) have negative interference with the standard model and pull the cross section down. The top-rho tends to increase
the cross section at low mass. The effect is so large that in fact it is basically impossible for the top-rho to be less than
300 GeV. For large top-rho mass, the contribution is negative. Note that even for a 500 GeV top-rho and light top-pion
and top-Higgs (black curve),  even though both the scalars and the vector separately want to lower the cross section,
their combined effect is a small positive contribution, at the $\lsim 5\%$ level.

\subsection{Same-sign tops}
\label{sec:tt}
One of the most pressing problems for $t$-channel explanations of the top asymmetry is that these models generically
produce an abundance of same-sign tops. Indeed the same $\tbar u X$ coupling explaining $\afb$ automatically allows for $uu\rightarrow tt$.
Even at the Tevatron, this could have been seen. At the LHC, although no same-sign top bound has been published yet,
any same-sign top excess would have lit up the inclusive search for same-sign lepton pairs. Thus it is impossible
to have a large contribution to same-sign tops.

To get a bound, we use the published Tevatron bounds. There are two CDF results. One~\cite{tt-cdf1},
which is model dependent, has a new particle $\phi$ which produces $tt$ through $t$-channel exchange. This particle
is assumed to be heavier than $m_t$ so there are also contributions from $u g \to \phi t \to t t \bar{u}$ and
$u \bar{u} \to \phi \phi \to t t \bar{u} \bar{u}$. The combined bound comes out to $\sigma_{tt} \lsim$ 800 fb.
The second CDF result~\cite{tt-cdf2} uses more data but looks exclusively for $tt$ states. The bound here is $\sigma_{tt} \lsim$ 500~fb.
We will use this second bound since it is stronger.

For the $\ttbar$ cross-section, we saw that interference between new physics and the standard model, and interference
among the new physics particles themselves conspired to have a small effect on the total cross section.
For $tt$ production, in contrast, the standard model contribution is tiny and comes only from weak processes. So
the interference between new physics and the standard model cannot help.
However, interference between different topcolor processes producing $tt$ could be negative. This is illustrated in Table~\ref{tab:tt_int}, where we list the contribution from interferences to the matrix element squared. In fact,
there is an exact cancellation between the top-Higgs and top-pion in the limit that their masses are degenerate and they
couple with the same strength.

\begin{table}[t]
  \begin{tabular}{|c|c|c|c|}
    \hline
    $|{\mathcal M}(u u \to t t)|^2$ & $\pi$ & $h_t$ & $\rho$\\
    \hline
    $\pi$   & +$\frac{g_{\pi }^4 m_t^4}{\left(m_{\pi }^2+m_t^2\right){}^2}$ &
    $-\frac{2 g_h^2 g_{\pi }^2 m_t^4}{\left(m_h^2+m_t^2\right) \left(m_{\pi }^2+m_t^2\right)}$ & 
+$\frac{16 g_{\rho }^4 m_t^4}{3 (m_\pi^2 + m_t^2) m_{\rho }^2}$\\
    \hline
    $h_t$   & &$\frac{g_h^4 m_t^4}{\left(m_h^2+m_t^2\right){}^2} $&
 $-\frac{16 g_h^2 g_{\rho }^2 m_t^4}{3 \left(m_h^2+m_t^2\right) m_{\rho }^2}$\\
    \hline
    $\rho$&  &  & 
+$\frac{32 g_{\pi }^2 g_{\rho }^2 m_t^4}{3 m_\rho^4}$\\
    \hline
  \end{tabular}
  \caption{\label{tab:tt_int}
Interference terms in same sign top production. To see the rough scaling properties
and signs, we have taken $s = 4 m_t^2, t =u=-m_t^2$ and $m_\rho \gg m_t$. The full matrix elements are given in Appendix \ref{app:tt}.}
\end{table}   

There are various ways to understand the top-pion/top-Higgs interference effect.
The cancellation can be seen by direct computation: top-pion exchange produces a $tt$ amplitude with
the same magnitude but opposite sign as top-Higgs exchange, due to the extra $i\gamma_5$ in the pseudoscalar interaction.
There is also a symmetry interpretation. If we write $\phi = \pi + i h$ as a complex scalar, then $\phi$ can
be thought of as carrying a chiral charge whose conservation forbids $uu\rightarrow tt$.
This mechanism was employed in \cite{Nelson:2011us} where a complex scalar was introduced directly for this purpose. Here we are observing
that the top-pion and top-Higgs automatically combine into this complex scalar. The symmetry is broken by the
top-pion/top-Higgs mass splitting and any difference in couplings. But there is still a large cancellation even if
the particles are separated by hundreds of GeV.

\begin{figure}
\psfrag{X}[]{$m_\pi$}
\psfrag{Y}[]{$\sigma_{tt}$ (fb)}
\psfrag{W}[t]{$p\bar{p} \to tt$}
\begin{center}
\includegraphics[width=0.45\textwidth]{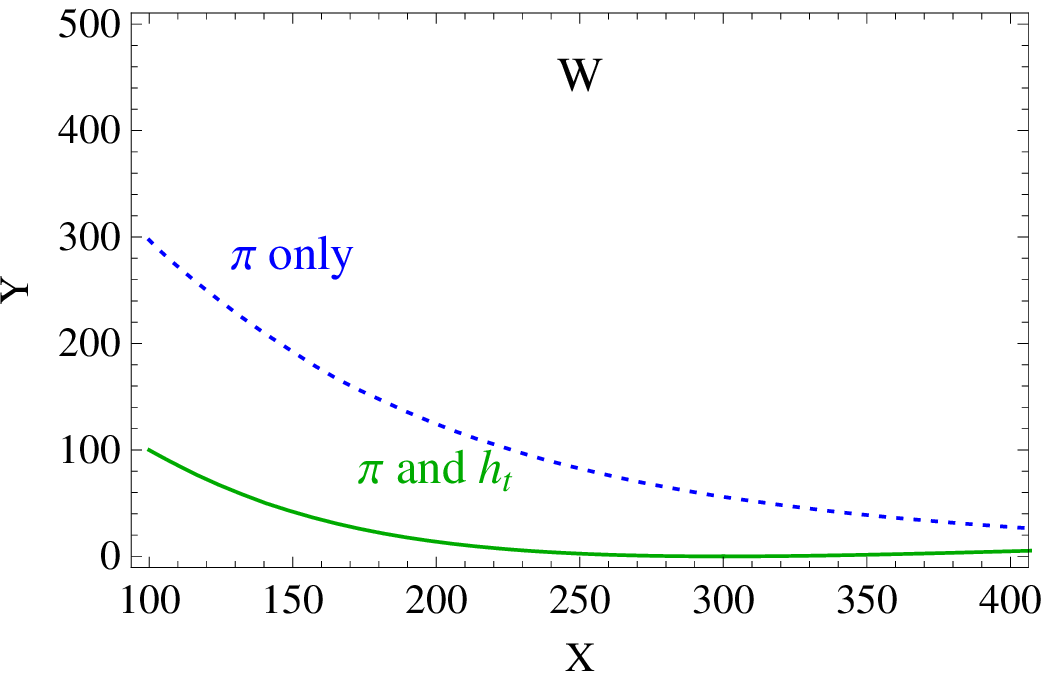}
\psfrag{Z}[t]{$~~~~~~p\bar{p} \to tt +tt\bar{u} + tt \bar{u}\bar{u}$}
\includegraphics[width=0.45\textwidth]{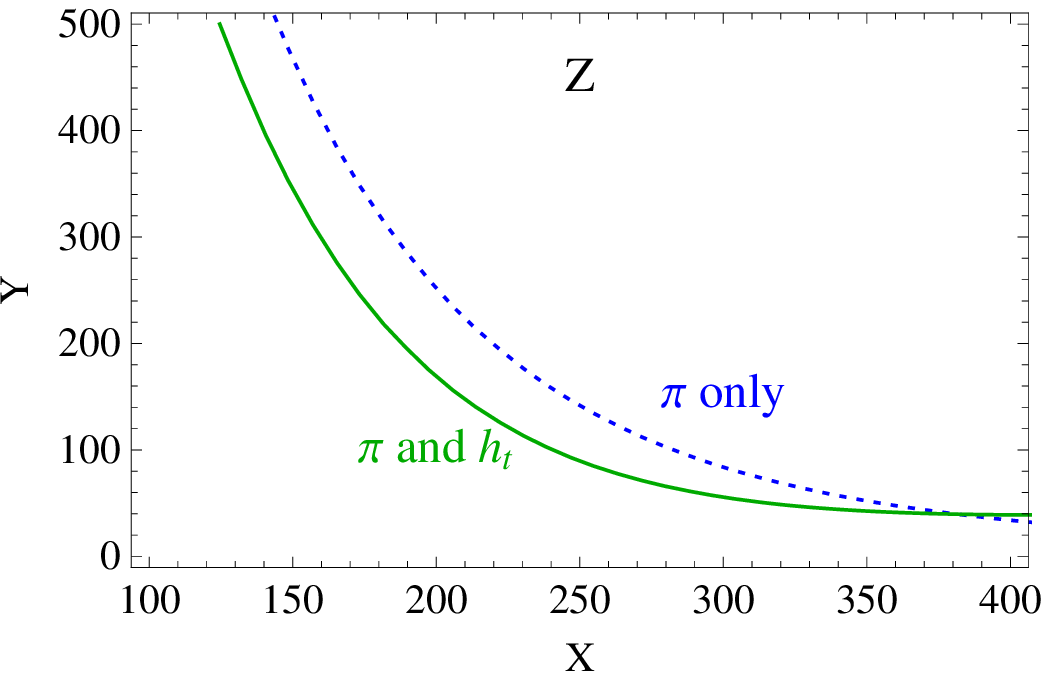}
\caption{Same-sign top production cross section at the Tevatron, as a function of top pion mass.
The left panel shows the cross section for $tt$ (and $\bar{t}\bar{t}$) production only, the right
panel also includes on-shell top-pion and top-Higgs decays (to same-sign $tt$ and $\bar{t}\bar{t}$).
In both plots, the dotted blue curve shows the top-pion alone (which is the same as for the top-Higgs). The
solid green curve is the cross section as a function of the top-pion mass when the top-Higgs is also present with fixed
mass of $m_{h_t}=300 \gev$.
 All couplings are set to 1.
\label{fig:SameTTsingle}
}
\end{center}
\end{figure}

Figure~\ref{fig:TTsingle} shows the same-sign top cross section for the top-pion alone, which is the same
as the cross section for the top Higgs alone, and also the cross section when both the top-pion and top-Higgs are present. The
interference substantially suppresses the same-sign top rate. The left panel
shows the rate for just $tt$ and $\bar{t}\bar{t}$ production. There is also a contribution to same-sign tops from the
top-rho and also from on-shell top-pion/top-Higgs pair production, or production in association with a top, with $\pi/h \to t u$.
 These contributions are included in the right panel. We have not included interference effects for these extra channels, although
when the top-pion and top-Higgs are close in mass, there will be destructive interference in these channels as well.

\subsection{Combined constraints}
\label{sec:comb}
Having discussed all the experimental constraints separately, we will now consider combining all the measurements together.
Estimates in topcolor based on the NJL model and scaled-up QCD give us some sense of what parameter range is natural.
Following these estimates, we consider $m_\pi \sim (100-400) \gev$, $m_h\sim (200-400) \gev$, $m_\rho \gsim 400 \gev$ and couplings
between 0 and 1.5.

First, suppose we just have the top-pion. Looking at Figure~\ref{fig:Afb}, we can see that we would need a coupling larger than $2$ and
a pion mass $< 150$ GeV  to explain the asymmetry. In this regime from Figure~\ref{fig:TTsingle}
we see that there would be too little $\ttbar$ produced, with around a $60\%$ decrease from the standard model as well as a
marginally unallowed production of same-sign tops, as we see in Figure~\ref{fig:SameTTsingle}. Note that for $m_\pi < m_t$, the
channels which include the top-pion decaying to $t\ubar$ do not contribute.

Next, let us include the top-Higgs. This helps with the $\ttbar$ cross section and with same-sign tops. However, it is still hard to get enough asymmetry. Finally, we add the top-rho. This additional $\rho$ contribution can add to the asymmetry without producing
much more $\ttbar$ or same-sign tops. Moreover, due to interference between the top-rho and the top-scalars, the combination of
the 3 particles is more than the sum of their parts. There is a large parameter space, including regions with light top-pions as
well as regions where the top-pions are heavier than $m_t$.

\begin{figure}
\begin{center}
\psfrag{X}[]{$g_\pi$}
\psfrag{Y}[]{$g_\rho$}
\psfrag{R}[]{Allowed}
\psfrag{A}[b]{{\white Too much same-sign top}}
\psfrag{B}[]{{Too much $\ttbar$}}
\psfrag{C}[]{{Too little $\ttbar$}}
\psfrag{S}[t]{Too little}
\psfrag{T}[t]{{$\afb$}}
\psfrag{D}[b]{{{$~m_\pi =150 \gev$}}}
\psfrag{E}[b]{{{$m_{h_t} =200 \gev$}}}
\psfrag{F}[b]{{{$m_\rho = 500 \gev$}}}
\includegraphics[width=0.45\textwidth]{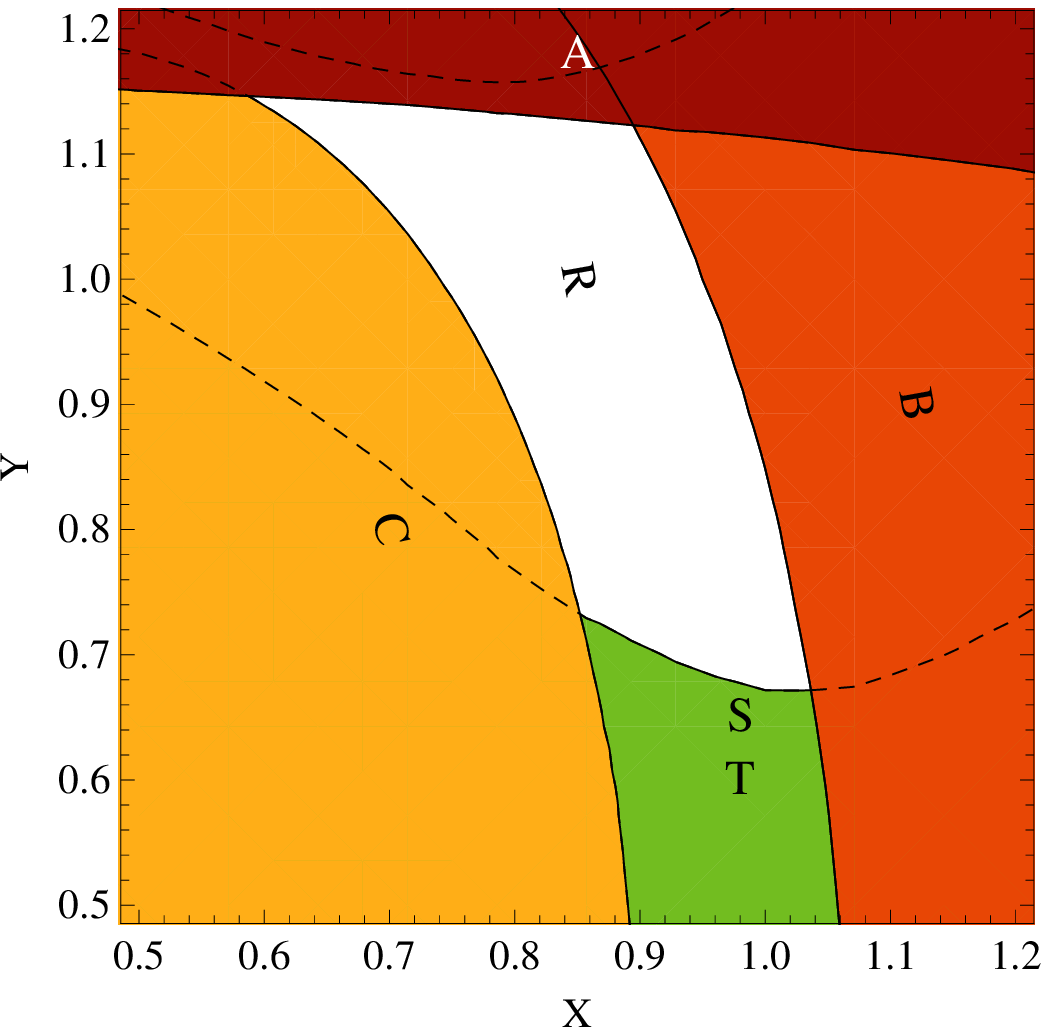}
\psfrag{D}[b]{{{$~m_{h_t} = 250 \gev$}}}
\psfrag{E}[b]{{{$m_\pi = 350 \gev$}}}
\psfrag{F}[b]{{{$m_\rho = 600 \gev$}}}
\psfrag{U}[t]{Too little $\afb$}
\includegraphics[width=0.45\textwidth]{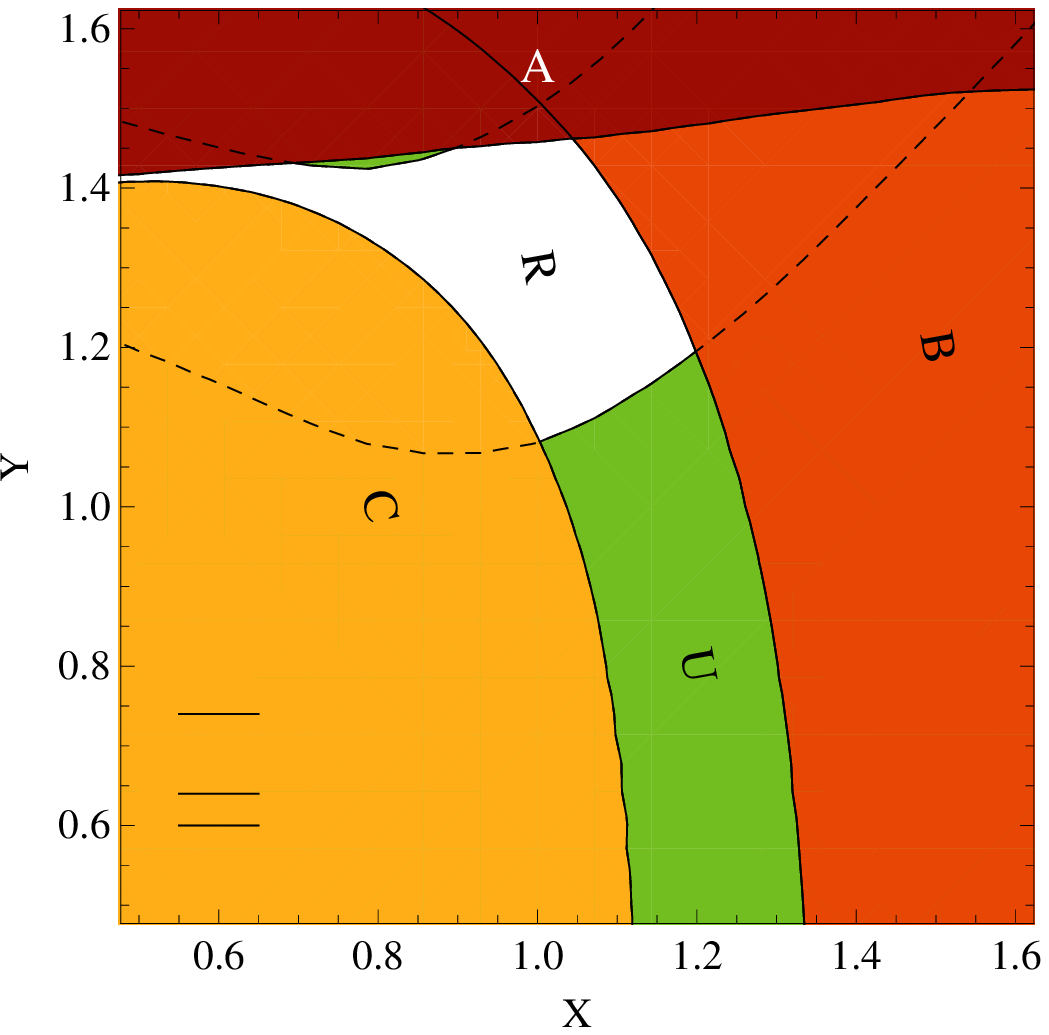}
\caption{Allowed region in coupling-constant space for two mass points. On the left is $m_\pi = 150$, $m_{h_t} = 200$, $m_\rho=500$ GeV
and on the right is  $m_\pi = 350$, $m_{h_t} = 250$, $m_\rho=600$ GeV. The band on top is excluded by
same-sign top cross sections at the Tevatron ($\sigma < 500$ fb),
 which includes on-shell pion decays to $tu$ in the right panel, but not
the left. The bounds from the $\ttbar$ cross section at the Tevatron are also shown. Note that
a large region of parameter space is ruled out by the predicted $\ttbar$ cross section being two low. 
The black lines outline the $\afb$ predictions, between 0.1 and 0.3. The central uncolored region is allowed.
\label{fig:scan}
}
\end{center}
\end{figure}

In Figure~\ref{fig:scan} we show the allowed parameter space for the $tu$ couplings $g_\pi$ and $g_\rho$ (see Eq.~\eqref{gpirho})
for two representative mass points.
On the left, we take $m_\pi = 150$, $m_{h_t} = 200$ and $m_\rho=500$ GeV. This has a light top-pion, below the top-mass. The top-pion can also be heavier than $m_t$, as shown on the right, where $m_\pi = 350$, $m_{h_t} = 250$ and $m_\rho = 600$ GeV. Note that for the second mass point, we have inverted the hierarchy between $m_\pi$ and $m_{h_t}$, introducing destructive interferences that helps ameliorate the same sigh top bounds.

\begin{figure}
\psfrag{X}{$m_\pi$}
\psfrag{Y}{$g_\pi$}
\psfrag{E}{Excluded}
\psfrag{A}{Allowed}
\psfrag{R}[b]{$\pi, h_t$ and $\rho$}
\psfrag{S}[t]{$\pi$ and $h_t$}
\begin{center}
\includegraphics[width=0.7\textwidth]{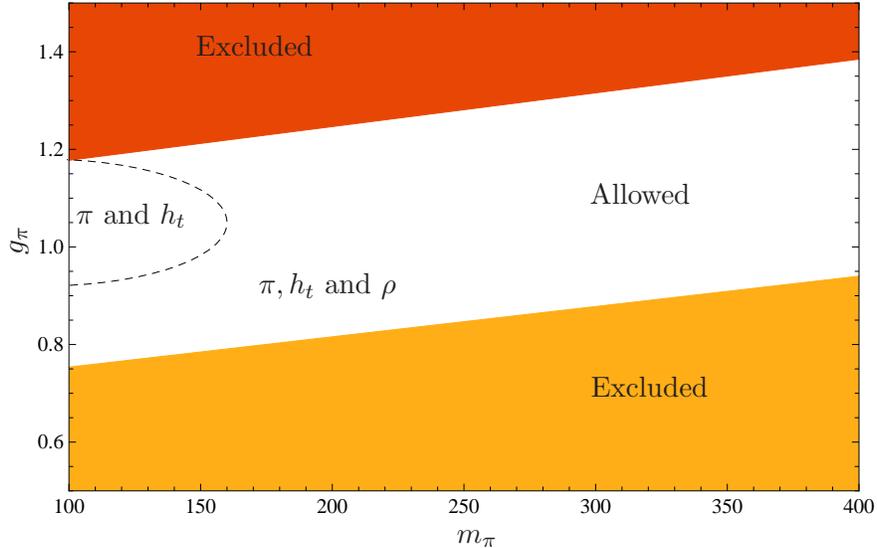}
\caption{Allowed region of top-pion mass and coupling, where any values of $g_\rho$, $m_\rho$ and $m_{h_t}$ are considered.
Allowed region satisfies the $\sigma_{\ttbar}$ and same-sign top constraints, while producing a top asymmetry consistent
with the measured value. The region on the left is allowed if the $\rho$ is removed from the spectrum.
\label{fig:pionmass}
}
\end{center}
\end{figure}

In Figure~\ref{fig:pionmass}, we show the allowed range of pion masses. To generate these curves, we allowed the top-Higgs
and top-rho masses to vary over $400 \gev < m_\rho < 800 \gev$ and $200 < m_h < 400$ and $0 < g_\rho < 1$. The region
with no top-rho ($g_\rho = 0$) is shown on the left side. Without a top-rho, the top-pion mass has to be very light.
A light top-pion is probably ruled out by direct searches, although the precise bound depends on how the
top-pion decays, which is somewhat model dependent. If we look for the region of top-pion and top-rho masses which
are allowed by the $\sigma_\ttbar$ and same-sign top bounds, practically any values of $m_\pi$ and $m_\rho$ are allowed
for some coupling strength and $m_h$ mass. In fact, taking $m_h=m_\pi$ removes the same-sign bound completely and
the $\sigma_\ttbar$ bound is controlled by destructive interference between the new physics and the standard model,
as we have discussed.. 
\section{Other signatures}
\label{sec:more}
We have shown the neutral top-pion, the top-Higgs and the neutral top-rho can combine to give a large top forward-backward asymmetry
with small effects on the total $\ttbar$ cross section and a not-yet-detectable number of same-sign tops produced. These new
particles  do not generate any new strong flavor constraints. Particles closely related to these, such as the charged top-pions
or the charged top-rhos, can contribute to observables like $Z\to b\bar{b}$. However, the charged top-pions could be heavy and the constraints would be weak. From the model-independent
point of view the main point is that there are no strong flavor constraints directly involving only the particles related to the $\afb$ explanation.

The top-mesons will have interesting signatures at the LHC depending on how they are produced and decay. Generically, we know
there should be a sizable coupling of these particles to $\ttbar$ and $t u$. If the top-pion is lighter than the top, it
can decay to $W b \ubar$ through an off-shell-top. Particles heavier than the top could decay directly to $t u$. If the
top-rho is heavy enough, it will decay to $W+\pi$ or $WW$ rather than $t u$.
The top mesons can be singly produced through gluon fusion through a top-loop. They can also be produced in association with a top,
or a $W$ or $Z$ or pair produced from $u\bar{u}$ through the exchange of a $t$-channel top. There are many possibilities.

A generic consequence of having particles coupling strongly to $tu$ is that there should be $t u$ resonances visable at the LHC.
For example, top-pion pair production would give a large number of  $\ttbar u\ubar$ events. A single top-pion could be
produced in association with a top through $g u \to u \to t \pi$. Thus there would also be $\ttbar u$ events. In general,
looking at events with tops and jets and looking at top-jet invariance mass distributions should discover or rule out this
model fairly early on at the LHC.

One more consequence of the couplings we have discussed is that if there is a large $\pi t u$ coupling and the top-pion is
lighter than the top, then we could have $t \to \pi u$. Indeed, for $m_\pi = 150 \gev$ with a coupling $g_\pi  =1$,
 the branching ratio to this channel would be 7\%, with the top  decaying 93\% of the time to $W b$. What this
looks like depends on how the top-pion decays. The top-pion might decay to $\bbbar$. This coupling is difficult to calculate
if the $b$-quark mass comes from non-perturbative effects (instantons) in the top-color sector. It does get a calculable
contribution from top/top-pion loops which make some branching fraction to $\bbbar$ inevitable.
If the $\bbbar$ coupling is small enough then the decay mode  $\pi \to c \cbar$  or $\pi \to u \ubar$ might dominate.
 Alternatively, the decay $\pi \to W u \bar{b}$ through an off-shell
top could dominate as well, depending on the couplings. In any case, the observed rate for the $\ttbar$ cross section
is based on a leptonic branching ratio taken from the standard model. So it is possible that that $\ttbar$ rate might
be slightly higher, with a slightly lower branching ratio to $Wb$ making the top production seem more consistent with
the standard model than it actually is.

Let us consider the situation in which $\pi \to c\bar{c}$. Then we would have $t \to u c\bar{c}$ 7\% of the time. This excess
of hadronically decaying tops could help explain an excess seen by CDF in events with two massive fat jets~\cite{pekka}. It might
also explain the recently observed excess in multijets near the top-mass seen by a different CDF group~\cite{Aaltonen:2011sg}. In fact,
this decay mode could also explain the excess in $W$+jets events seen by CDF \cite{Aaltonen:2011mk}. For example, this excees could be produced
by a top-rho decaying to $W + \pi$, with $\pi \to c\bar{c}$. If $\pi$ decays to $\bbbar$ dominantly, then it is hard to explain
the CDF excess since the peak does not seem to be rich in $b$'s. It is unclear without futher model
details and more calculations whether the rate for production of these modes is consistent with the size of the excess.

If the top-pions are heavier than $m_t$, they cannot be the explanation of the CDF $W$+jets bump. However, it
is still possible that the bump might be due to particles in the technicolor sector, as proposed in~\cite{Eichten:2011sh}. These
particles are also present in topcolor-assisted technicolor, but we have nothing new to say about them from previous work.

\section{Conclusion}
\label{sec:conc}
The measurements of the top-quark forward-backward asymmetry at the Tevatron
suggest new physics may be showing up strongly coupled to the top. That new physics would
show up here is not surprising to the many theorists who have had to contend with
the large mass hierarchy between the top mass and the other fermions. Indeed, top-condensation and
its realization in topcolor~\cite{Hill:1991at} was suggested
more than 2 decades ago as a natural combination of dynamical symmetry breaking with
an explanation of the large top quark mass. Topcolor even predates the discovery of the top quark itself.
When topcolor is combined with technicolor, the resulting framework, topcolor-assisted technicolor (TC2),
provides a solution of the hierarchy problem and the large top quark mass without some of the strongest
flavor-changing neutral current problems of extended technicolor itself. The combined framework
explains the origin of mass as a team effort: the top mass comes from top-color, the $W$ and $Z$ masses
come mostly from technicolor, the $b$-quark mass comes from instantons, and the light quark masses come from extended technicolor.
Taking this one step further, most of the visible matter in the universe is baryons,
which get their masses from QCD, neutrino masses can come from the seesaw mechanism associated with additional new
physics at a high scale, and dark matter might come from a totally decoupled sector, such as from axions.

In this paper, we have shown that TC2 also contains all ingredients for producing the large $A_{\text{FB}}^{t\bar t}$ anomaly
without violating other constraints. The spectrum contains three relevant particles, the neutral top-pion, the top-Higgs and
the neutral top-rho. All of these can have large top-up couplings due to mixing of the right-handed top with the lighter
up-type right-handed quarks. This mixing can be naturally large in TC2. Thus, there is automatically an excess
in the top-quark forward-backward asymmetry over the standard model due to $t$-channel exchange of these particles. Intriguingly
there is constructive interference among the particles in producing a large $\afb$, but destructive interference affecting
the overall $\ttbar$ cross section. Same-sign top production is a signature of generic points in parameter space of this
model. If the top-Higgs and neutral top-pion are close in mass, the current experimental bound on
same-sign top production can be avoided without much tuning. Nevertheless, without additional symmetries, it
seems impossible to avoid a same-sign top-signal, particularly from on-shell pion or top-Higgs decays, which should
be visible at the LHC early on. This would be a clear signature of this type of explanation of the $\afb$ excess.

There are no devastating flavor problems in TC2, even with large mixing. Of course, TC2 does have difficulty with precision
electroweak, but there is still no reliable way to calculate $S$ and $T$ and new particles can also contribute. Thus it
is impossible to rule out the model based on precision electroweak alone. Flavor-changing neutral currents are not bad in
TC2 because topcolor only couples to top, at leading order, so it naturally falls into next-to-minimal flavor violation \cite{Agashe:2005hk}.
The light quark masses are generated with extended technicolor. However, because ETC is no longer responsible for the bottom
or top masses, the tuning required is not very severe. Constraints such as the $Z\to \bbbar$ rate, $R_b$, are dangerous,
but not directly related to the particles responsible for explaining $\afb$. To the extent that flavor physics is under
control in previous incarnations of TC2, they are under control here.

Top-color assisted technicolor has a number of signals that will show up early on at the LHC. The top-pions and other
top-mesons should be produced in abundance. They can be singly produced through gluon fusion, as well as pair produced
through $t$-channel top-exchange or produced in association with tops through $ug$ initial states.
Top-pions can also be produced from decays of heavier top-particles. With large $tu$ couplings necessary to explain $\afb$,
there should be an abundance of events with multiple tops and multiple jets. Looking for a resonance peak in the
top-jet invariant mass would be a clean test for this model.
Thus, forthcoming results from the LHC should soon 
reveal whether top-condensation can be the explanation to the large top forward-backward
asymmetry observed at the Tevatron.

\vspace{0.5cm}
{\bf{Note Added}}:
As this manuscript was being finalized, the CMS collaboration released a same sign top bound from early LHC running~\cite{Collaboration:2011dk}. This bound is somewhat stronger than the Tevatron bounds. Topcolor-assisted technicolor can still explain
$\afb$ and be consistent with this observation,
but the parameter space is more constrained: the top-Higgs and neutral top-pion should be closer in mass (if they are
degenerate with the same coupling strength, same-sign top production is absent) and the top-rho should also be on the heavier
side, which is anyway consistent with some expectations from topcolor~\cite{Chivukula:1990bc}.


\section*{Acknowledgments}
  We thank Sekhar Chivukula, Howard Georgi, Sunghoon Jung, Yasunori Nomura and James Wells for helpful discussions and Daniel Whiteson for helping us
understand the CDF same-sign top results. This work is supported by the NSF under grants PHY-0855591 
and PHY-0804450 and the DOE under grant DE-SC003916.
\appendix

\section{$\ttbar$ interference effects}
\label{app:ttbar}
In this appendix, we give formulas for the partonic matrix elements for $\ttbar$ pair production. These formulas
are useful for seeing the signs and strengths of various interference effects.

\begin{equation}
\M\M^*=\M_{gg}+\M_{\pi\pi}+\M_{h_th_t}+\M_{\rho\rho}+\M_{g\pi}+\M_{gh_t}+\M_{g\rho}+\M_{\pi h_t}+\M_{\pi\rho}+\M_{h_t\rho},
\end{equation} 
where $\M_{ii}$ ($i=g,\pi,h_t,\rho$) denote the squared terms for single diagrams and $\M_{ij}$ $(i\neq j)$ denote the interferences. Then we have
\begin{eqnarray}
\M_{gg}&=&\frac{4 g_s^4 \left(2 m_t^4-4 t m_t^2+s^2+2 s t+2 t^2\right)}{9 \
s^2},\\
\M_{\pi\pi}&=&\frac{g_{\pi }^4 \left(m_t^2-t\right){}^2}{4 \
\left(t-m_{\pi }^2\right){}^2},\\
\M_{h_th_t}&=&\frac{\left(m_t^2-t\right){}^2 g_{h_t}^4}{4 \
\left(t-m_{h_t}^2\right){}^2},\\
\M_{\rho\rho}&=&\frac{g_{\rho }^4 \left(4 s m_t^4 m_{\rho }^2+4 m_{\rho \
}^4 \left(-m_t^2+s+t\right){}^2+\left(m_t^2-t\right){}^2 m_t^4\right)}{4 \
m_{\rho }^4 \left(t-m_{\rho }^2\right){}^2},
\end{eqnarray}
and
\begin{eqnarray}
\M_{g\pi}&=&\frac{4 g_{\pi }^2 g_s^2 \left(m_t^2 (s-2 \
t)+m_t^4+t^2\right)}{9 s \left(t-m_{\pi }^2\right)},\\
\M_{gh_t}&=&\frac{4 g_s^2 g_{h_t}^2 \left(m_t^2 (s-2 \
t)+m_t^4+t^2\right)}{9 s \left(t-m_{h_t}^2\right)},\\
\M_{g\rho}&=&\frac{4 g_s^2 g_{\rho }^2 \left(m_t^2 \left(m_t^2 (s-2 \
t)+m_t^4+t^2\right)+2 m_{\rho }^2 \left(-m_t^2 (s+2 \
t)+m_t^4+(s+t)^2\right)\right)}{9 s m_{\rho }^2 \left(t-m_{\rho }^2\right)},\
\\
\M_{\pi h_t}&=&\frac{g_{\pi }^2 \left(m_t^2-t\right){}^2 g_{h_t}^2}{2 \
\left(t-m_{\pi }^2\right) \left(t-m_{h_t}^2\right)},\\
\M_{\pi\rho}&=&\frac{g_{\pi }^2 g_{\rho }^2 m_t^2 \left(2 s m_{\rho \
}^2+\left(m_t^2-t\right){}^2\right)}{2 \left(t-m_{\pi }^2\right) m_{\rho }^2 \
\left(t-m_{\rho }^2\right)},\\
\M_{h_t\rho}&=&\frac{g_{\rho }^2 m_t^2 g_{h_t}^2 \left(2 s m_{\rho \
}^2+\left(m_t^2-t\right){}^2\right)}{2 m_{\rho }^2 \left(t-m_{h_t}^2\right) \
\left(t-m_{\rho }^2\right)}.
\end{eqnarray}

\section{$tt$ interference effects}
\label{app:tt}
In this appendix, we give formulas for the partonic matrix elements for same-sign top production. 
The matrix element squared for $uu\rightarrow tt$ is written as
\begin{equation}
\M\M^*=\M_{\pi\pi}+\M_{h_th_t}+\M_{\rho\rho}+\M_{\pi h_t}+\M_{\pi\rho}+\M_{h_t\rho},
\end{equation} 
where
\begin{eqnarray}
\M_{\pi\pi}&=&\frac{g_{\pi }^4 \left(m_t^2-t\right){}^2}{4 \
\left(t-m_{\pi }^2\right){}^2}-\frac{g_{\pi }^4 \left(m_t^4-2 t m_t^2+t \
(s+t)\right)}{12 \left(t-m_{\pi }^2\right) \left(u-m_{\pi }^2\right)},\\
\M_{h_th_t}&=&\frac{\left(m_t^2-t\right){}^2 g_{h_t}^4}{4 \
\left(t-m_{h_t}^2\right){}^2}-\frac{g_{h_t}^4 \left(m_t^4-2 t m_t^2+t \
(s+t)\right)}{12 \left(t-m_{h_t}^2\right) \left(u-m_{h_t}^2\right)},\\
\M_{\rho\rho}&=&\frac{g_{\rho }^4 \left(4 s m_t^4 m_{\rho }^2+4 s m_{\rho \
}^4 \left(s-2 m_t^2\right)+\left(m_t^2-t\right){}^2 m_t^4\right)}{4 m_{\rho \
}^4 \left(t-m_{\rho }^2\right){}^2}+\nonumber\\
&&-\frac{g_{\rho }^4 \left(-4 s m_t^4 \
m_{\rho }^2+4 s m_{\rho }^4 \left(2 m_t^2-s\right)+m_t^4 \left(m_t^4-2 t \
m_t^2+t (s+t)\right)\right)}{12 m_{\rho }^4 \left(m_{\rho }^2-t\right) \
\left(m_{\rho }^2-u\right)},
\end{eqnarray}
and
\begin{eqnarray}
\M_{\pi h_t}&=&\frac{g_{\pi }^2 g_{h_t}^2 \left(m_t^4-2 t m_t^2+t \
(s+t)\right)}{6 \left(t-m_{\pi }^2\right) \
\left(u-m_{h_t}^2\right)}-\frac{g_{\pi }^2 \left(m_t^2-t\right){}^2 \
g_{h_t}^2}{2 \left(t-m_{\pi }^2\right) \left(t-m_{h_t}^2\right)},\\
\M_{\pi\rho}&=&\frac{g_{\pi }^2 g_{\rho }^2 m_t^2 \left(2 s m_{\rho \
}^2+\left(m_t^2-t\right){}^2\right)}{2 \left(t-m_{\pi }^2\right) m_{\rho }^2 \
\left(t-m_{\rho }^2\right)}-\frac{g_{\pi }^2 g_{\rho }^2 m_t^2 \left(-2 s \
m_{\rho }^2+m_t^4-2 t m_t^2+t (s+t)\right)}{6 \left(t-m_{\pi }^2\right) \
m_{\rho }^2 \left(u-m_{\rho }^2\right)},\\
\M_{h_t\rho}&=&\frac{g_{\rho }^2 m_t^2 g_{h_t}^2 \left(-2 s m_{\rho \
}^2+m_t^4-2 t m_t^2+t (s+t)\right)}{6 m_{\rho }^2 \left(t-m_{\rho }^2\right) \
\left(u-m_{h_t}^2\right)}-\frac{g_{\rho }^2 m_t^2 g_{h_t}^2 \left(2 s m_{\rho \
}^2+\left(m_t^2-t\right){}^2\right)}{2 m_{\rho }^2 \left(t-m_{h_t}^2\right) \
\left(t-m_{\rho }^2\right)}.
\end{eqnarray}


\end{document}
